\pdfoutput=1
\UseRawInputEncoding
\documentclass[aps,preprint,amsmath,amssymb,amsfonts,nofootinbib]{revtex4-1}
\usepackage{epsfig}
\usepackage{dcolumn}
\usepackage{bm}
\usepackage{amsthm}
\usepackage{amsmath}
\usepackage{color}

\newcommand{\beq}{\begin{equation}}
\newcommand{\eeq}{\end{equation}}
\newcommand{\bea}{\begin{eqnarray}}
\newcommand{\eea}{\end{eqnarray}}

\begin{document}
\setlength{\parskip}{0pt}

\title{A gauge theory for the 3+1 dimensional incompressible Euler equations}
\author{Christopher Eling}
\email{cteling@gmail.com}
\noaffiliation

\begin{abstract}

We show that the incompressible Euler equations in three spatial dimensions can be expressed in terms of an abelian gauge theory with a topological BF term. A crucial part of the theory is a 3-form field strength, which is dual to a material invariant local helicity in the fluid. In one version of the theory, there is an additional 2-form field strength, with the magnetic field corresponding to fluid vorticity and the electric field identified with the cross-product of the velocity and the vorticity. In the second version, the 2-form field strength is instead expressed in terms of Clebsch scalars.  We discuss the theory in the presence of the boundary and argue that edge modes may be present in the dual description of fluid flows with a boundary.

\end{abstract}

\maketitle

\section{Introduction}

The role of topology in the study of fluid dynamics was first realized in 19th century, when Helmholtz, Lord Kelvin, and others realized that vortex structures are invariant quantities carried along by an ideal fluid flow.  For example, the degree to which vortex tubes are tangled and knotted is a material invariant and persists in the flow.  Furthermore, the velocity circulation around a loop moving with the flow depends on whether the curve encloses a simply or multiply connected domain.  For further applications of topological methods to fluids, see \cite{Arnold}.

In addition to these well-known features of fluid flows, in recent years several non-trivial connections have emerged between the properties of waves in fluid systems and the topological phases of matter \cite{Delplace, Souslov, Tauber1, Perrot, Tauber2, Grad, Liu:2020ksx,Liu:2020abb, Venaille, Green:2020uye, Tauber3}.  One of the most striking results is that the shallow water wave modes localized near boundaries (eg, coasts, or even the equator) are equivalent to edge modes in topological phases of matter such as the quantum Hall state.  At long distances, quantum Hall states are described by Chern-Simons theory, which is a topological gauge invariant field theory.  For a review, see \cite{Dunne:1998qy,Tong:2016kpv}.  This connection between water waves and gauge theory was made explicit in \cite{Tong:2022gpg}, where it was shown that the shallow water equations can be encoded into a 2+1 dimensional gauge theory defined in terms of a U(1) gauge connection $A_\mu$.  The action for the linearized theory includes a Chern-Simons term that captures both the Poincare waves in the bulk and the boundary Kelvin waves localized near coasts.

Is it possible that the dual gauge theory representation can be extended to other fluid systems? In \cite{Eling:2023iyx} we proposed that the incompressible Euler equations can also be expressed in terms of a 2+1 dimensional gauge theory.  The 2d vorticity equation is identified with the Bianchi identity, which is a topological conservation law.  Hence, the scalar vorticity of the flow is mapped into the magnetic field, while the electric field is identified with the product of the vorticity and the gradient of the stream function.  We then showed that the field equations arising from the variation of a certain gauge invariant action with an explicit Chern-Simons term reproduce the incompressible 2d Euler equation.

A natural question is whether 3+1 dimensional incompressible flows can also be mapped into a dual gauge theory, or is our result just a special feature of physics in three spacetime dimensions? In this paper we argue that the 3+1 dimensional Euler equations can be encoded into a gauge theory with a 2-form potential $P_{\mu \nu}$.  As before, the idea is that the equations expressing the material invariance of certain quantities are Bianchi identities of the gauge theory. In this case, the Bianchi identity associated with gauge transformations of $P_{\mu \nu}$ reflects the material invariance of a local helicity variable in the fluid description.

As a first step, we assume the 3+1 dimensional theory still contains a 1-form gauge potential $A_\mu$ as a fundamental variable. In one higher dimension the Bianchi identity is naturally identified with the 3d vorticity equation. We show that the field equations produced by varying a gauge invariant action with a topological ``BF'' term are the incompressible 3d Euler equations. However, these field equations also place a restriction on the local helicity.

To remedy this issue, we introduce Clebsch scalars as fundamental variables describing the vorticity sector of the theory, following \cite{Tong:2022gpg}. In this case, the action involves an auxiliary connection and a type of BF term we will refer to as Clebsch BF.  The resulting field equations yield the Euler equations and the material transport of the Clebsch scalars without restriction on the local helicity.

The structure of this paper is as follows.  In Section 2 we review the construction of the gauge theory in 2+1 dimensions and describe how to generalize the approach to the 3+1 dimensional case.  Along the way we review the concept of vortex momentum in a flow and how this leads to the material invariance (or Lie transport) of local helicity.  In Section 3 we construct the action principles for the theory.  In Section 4 we discuss the behavior of the theory in the presence of a boundary.  When a boundary is present, the gauge invariance of the theory is restricted, which leads to physical edge modes degrees of freedom.  We compute the quasi-local Noether charges for the theory and derive the charge algebra associated with the edge modes.  We conclude the paper with a discussion of open questions, including some speculative ideas on how the gauge theory picture can be extended to describe the viscous term in the Navier-Stokes equations.

\section{Gauge theory approach to incompressible fluids}

\subsection{From two to three dimensions}

In 2+1 dimensions, the gauge invariant action principle takes the form \cite{Eling:2023iyx}
\begin{align}
S_{\rm gauge} = \int dt d^2 x \left(\frac{E_i E^i}{2B} - pB -  L_{\rm CS} \right) \label{2daction}.
\end{align}
The action is expressed in terms of a gauge connection 1-form $A_\mu$ and field strength $F_{\mu \nu} = 2 \partial_{[\mu} A_{\nu]}$.  The electric field $E_i = F_{0i}$ and magnetic field $B = \epsilon^{ij} \partial_i A_j$.  $p$ is an auxiliary function we will identify with the fluid pressure.  The topological Chern-Simons term has the form
\begin{align}
L_{\rm CS} = \frac{1}{2} \epsilon^{\mu \nu \rho} A_\mu \partial_\nu A_\rho.
\end{align}
This term is independent of the metric and invariant under gauge transformations $A_\mu \rightarrow A_\mu + \partial_\mu \lambda$, up to a total derivative term.

Variation of this action with respect to $A_0$ yields the Gauss law constraint
\begin{align}
\partial_i \left(\frac{E^i}{B} \right) - B = 0  \label{Gauss},
\end{align}
while variation with respect to $A_i$ gives
\begin{align}
-\partial_t \left(\frac{E^i}{B} \right) - \epsilon^{ij} \partial_j \left( \frac{E^2}{2B^2} \right) - \epsilon^{ij} \partial_j p + \epsilon^{ij} E_j = 0.  \label{Euler1}
\end{align}
Furthermore, in a 2+1 dimensional gauge theory, the Bianchi identity implies
\begin{align}
\epsilon^{\mu \nu \rho} \nabla_\mu F_{\nu \rho} = \epsilon^{\mu \nu \rho} \partial_\mu \partial_\nu A_\rho = 0,
\end{align}
which can be rearranged into the form of the conservation of a current
\begin{align}
\partial_\mu \left(\epsilon^{\mu \nu \rho} F_{\nu \rho} \right) = 0.
\end{align}
This yields a Faraday law:
\begin{align}
\partial_t B - \epsilon^{ij}\partial_i E_j = 0. \label{Bianchi}
\end{align}
This set of equations is equivalent to the incompressible Euler system with the identifications
\begin{align}
B &= \omega \nonumber \\
E_i &= B \epsilon_{ij} v^j \label{Electricid},
\end{align}
where $\omega$ is the ultimately the fluid vorticity scalar and $v^i$ is ultimately the fluid velocity.  We assume the fluid velocity is divergence free, so that $\partial_i v^i = 0$, which implies $v^i= \epsilon^{ij} \partial_j \psi$ in terms of the stream function $\psi$.

A priori, the velocity and $\omega$ fields are independent variables.  The Gauss law (\ref{Gauss}) relates $\omega$ to the 2d curl of $v^i$ as expected
\begin{align}
\omega = \epsilon^{ij} \partial_i v_j.
\end{align}
The Bianchi identity (\ref{Bianchi}) becomes the ideal vorticity equation
\begin{align}
\partial_t \omega + v^i \partial_i \omega = 0.
\end{align}
Finally (\ref{Euler1}) is the Euler equation
\begin{align}
\partial_t v^i + v^j \partial_j v^i + \partial^i p = 0.
\end{align}

In the 3+1 dimensional case we can start by attempting a similar procedure, first identifying the Bianchi identity equation with the 3d vorticity equation.  In 3+1 dimensions the Bianchi identity is now expressed as the conservation of a 2-form current
\begin{align}
\partial_\mu \left(\epsilon^{\mu \nu \rho \sigma} F_{\rho \sigma} \right) = 0.
\end{align}
This yields two of the familiar Maxwell equations:
\begin{align}
\vec{\nabla} \cdot \vec{B} = 0 \\
\partial_t \vec{B} + \vec{\nabla} \times \vec{E} = 0,
\end{align}
where the magnetic field $B^i = -\frac{1}{2} \epsilon^{ijk} F_{jk}$ and the electric field $E_i = F_{0i}$.

A natural guess is to identify the magnetic field vector with the vorticity vector:
\begin{align}
\vec{B} = \vec{\omega}. \label{Bfield}
\end{align}
Then the first equation states (correctly) that the divergence of $\vec{\omega}$ is zero
\begin{align}
\vec{\nabla} \cdot \vec{\omega} = 0.
\end{align}
In three dimensions, vorticity is a vector and the vorticity equation is
\begin{align}
\partial_t \vec{\omega} = \vec{\nabla} \times (\vec{v} \times \vec{\omega}),
\end{align}
or in index notation for a vector vorticity
\begin{align}
\partial_t \omega^i + v^j \partial_j \omega^i - \omega^j \partial_j v^i = 0 \label{vorticityeqn1}.
\end{align}
Therefore, matching the vorticity equation with the Faraday law equation implies that the electric field should be identified as
\begin{align}
\vec{E} = -\vec{v} \times \vec{B} = -\vec{v} \times \vec{\omega} \label{Efield1},
\end{align}
which is the generalization of (\ref{Electricid}) to 3d.  Note that we are assuming that $\vec{v}$ is a divergence free vector; this condition does not appear as a separate topological conservation law\footnote{In 2+1 dimensions, the continuity equation for a compressible fluid with variable density $\rho$ can be expressed as a Bianchi identity for a 1-form gauge connection, with magnetic field $B=\rho$. See \cite{Tong:2022gpg,Nastase:2023rou}}.  In the 2d case discussed above the Euler equation and its active scalar generalizations \cite{Eling:2023iyx} are formulated directly in terms in the stream function, which can be considered a fundamental variable. In 3d, one could generalize to a ``stream vector'' $\vec{\psi}$ via $v^i = \epsilon^{ijk} \partial_j \psi_k$.

If a topological term is a crucial part of the action for incompressible fluids in general dimension, then a major obstacle in proceeding further is that a Chern-Simons term does not exist in 3+1 dimensions.  Instead, we will consider topological terms of the ``BF'' type, which can be defined in general dimension.
In 3+1 dimensions, terms of the BF type involve an additional 2-form gauge potential $P_{\mu \nu}$ (we will use $P_{\mu \nu}$ here to avoid confusion with magnetic field $\vec{B}$) and the topological Lagrangian has the form
\begin{align}
L_{\rm BF} = \epsilon^{\mu \nu \rho \sigma}  P_{\mu \nu} F_{\rho \sigma},
\end{align}
where $F_{\mu \nu}$ is still the field strength of $A_\mu$.  This action term is explicitly invariant under the 1-form gauge transformations of $A_\mu$, and is also invariant under the 2-form gauge transformations
\begin{align}
P_{\mu \nu} \rightarrow P_{\mu \nu} + \partial_{[\mu} \lambda_{\nu]}. \label{2formgauge}
\end{align}
up to a total derivative term.

In two spatial dimensions the 1-form gauge transformations have previously been linked to the area preserving diffeomorphism invariance of the fluid theory when viewed from the Lagrangian perspective \cite{Susskind:2001fb,Sheikh-Jabbari:2023eba}.  Lagrangian coordinates $\eta_i$ parametrize the locations of each fluid parcel at a given time.  In 2d, area preserving transformations of the Lagrangian coordinates take the form
\begin{align}
\eta’_i = \eta_i + \epsilon_i{}^j \partial_j \lambda(\eta). \label{areadiffeos}
\end{align}
One can then consider small perturbations around an equilibrium solution where the Lagrangian coordinates are equal to Eulerian space coordinates $x_i$.  If the perturbations are parametrized with the 1-form field $A_i$,
\begin{align}
x_i = \eta_i + \epsilon_i{}^j A_j(\eta), \label{mapgauge}
\end{align}
then to linear order the area preserving transformation is equivalent to a U(1) gauge transformation of $A_i$
\begin{align}
A_i \rightarrow A_i + \partial_i \lambda.
\end{align}
Note that in this mapping to a gauge theory one has fixed the temporal gauge $A_0 = 0$.  Hence, the area preserving diffeomorphisms are equivalent to the remaining residual gauge transformations in the linearized regime.

In 3d, the volume preserving transformations are a generalization of (\ref{areadiffeos})
\begin{align}
\eta’_i = \eta_i + \epsilon_i{}^{jk} \partial_j \lambda_k(\eta). \label{volumediffeos}
\end{align}
Similarly, one can introduce a 2-form field to parametrize perturbations about the equilibrium configuration
\begin{align}
x_i = \eta_i + \epsilon_i{}^{jk} P_{jk}(\eta). \label{mapgauge2}
\end{align}
then at the linear level, the volume preserving diffeomorphisms act as the 2-form residual gauge transformations in (\ref{2formgauge})\footnote{In this case the mapping is to a gauge theory with the gauge fixing $P_{0i} = 0$}.  This heuristic argument hints that 2-form gauge transformations should play a role in the mapping to 3d fluid flows.

The field strength associated with $P_{\mu \nu}$ is the 3-form field $H_{\mu \nu \rho}$
\begin{align}
H_{\mu \nu \rho}  =  \partial_{[\mu} B_{\nu \rho]}.
\end{align}
In this theory there is another Bianchi identity involving $H_{\mu \nu \rho}$:
\begin{align}
\partial_{[\mu} H_{\nu \rho \sigma]} = 0.
\end{align}
This Bianchi identity equation can be expressed as conservation law of a current density
\begin{align}
\partial_\mu \left(\epsilon^{\mu \nu \rho \sigma} H_{ \nu \rho \sigma} \right) = 0,
\end{align}
or, equivalently,
\begin{align}
\partial_t H + \partial_i H^i = 0,
\end{align}
where $H = \epsilon^{ijk} H_{ijk}$ and $H^i = -3 \epsilon^{ijk} H_{0jk}$.

The question is how to identify the field strength $H$ with a fluid quantity.  An obvious choice is the fluid helicity density $h$, which is a scalar built out of a 3-form via the Hodge dual (ie exterior product of 1-form velocity and 2-form vorticity),
\begin{align}
 h = \vec{v} \cdot \vec{\omega},~   h_{ijk} = v_{[i} \omega_{jk]}.
\end{align}
Total helicity $\int h ~d^3 x$ is a famously a global conserved quantity in flows after imposing the Euler and vorticity equations \cite{Moffatt}.

However, what we'd really like is a Lagrangian material invariant quantity, analogous to the vorticity scalar in 2d.  In terms of the Lagrangian comoving derivative
\begin{align}
D_t \omega = \partial_t + v^i \partial_i \omega = 0.
\end{align}
On a general curved manifold this equation can be expressed in terms of the Lie derivative along $v$, leading to the more generic notion of ``Lie transport'' \cite{BesseFrisch}:
\begin{align}
\partial_t \omega + {\cal L}_v \omega = 0.
\end{align}
Similarly,  in 3d, the vorticity equation (\ref{vorticityeqn1}) can be expressed as Lie transport of $\vec{\omega}$:
\begin{align}
\partial_t \vec{\omega} + {\cal L}_v \vec{\omega} = 0.
\end{align}
Note that vorticity expressed as 2-form $\omega_{ij}$ is also Lie-transported.

The material invariance of the vorticity follows from the invariance of the ideal fluid theory under area/volume preserving diffeomorphisms (in an integrated form this is the Kelvin circulation theorem).  The philosophy is that in the gauge theory description locally conserved quantities correspond to the topological conservation laws constructed via the Bianchi identities. It turns out there is a type of local helicity constructed from ``vortex momentum density'' that is a material invariant, as we describe next.

\subsection{Vortex momentum and local helicity}

We start by considering the velocity field to be a 1-form on a general manifold.  In terms of the Lie derivative on a 1-form, the Euler equation has the form
\begin{align}
\partial_t v_i + {\cal L}_v v_i = \partial_i \left(\frac{1}{2} v^2 - p\right).
\end{align}
Hence, velocity is almost Lie transported, up to an exact 1-form.  Now introduce a scalar $\ell$ such that
\begin{align}
\partial_t \ell + {\cal L}_v \ell = \frac{1}{2} v^2 - p.
\end{align}
It is straightforward to show that field $u_i = v_i - \partial_i \ell$ is Lie-transported along the flow
\begin{align}
\partial_t u_i + {\cal L}_v u_i = 0.
\end{align}
The velocity potential $\ell$ was first introduced by mathematician Heinrich Martin Weber in the 19th century \cite{Weber} and the importance of the Lie transport of $u$ was recognized by Kuz’min and Oseledets in the 1980's \cite{Kuzmin, Oseledets}. $u$ can be interpreted a vortex momentum density \cite{Eyink} produced by a large number of infinitesimal vortex rings or vortex dipoles.  Each individual dipole has a linear momentum and the distribution of a large number of these dipoles can be thought of as producing $u_i$. Another related way of interpreting $u$ is as the density of the impulse $I = \int \vec{r} \times \vec{u} ~d^3 x$ in a region of the fluid \cite{Russo}

The vortex momentum helicity 3-form $u_{[i} \omega_{jk]}= v_{[i} \omega_{jk]} - \partial_{[i} \ell ~\omega_{jk]}$ is Lie transported because the product of two Lie transported quantities is also Lie transported.  Moreover, the dual scalar helicity density $\sigma = u_i \omega^i$ is also a Lagrangian invariant
\begin{align}
D_t \sigma = \partial_t \sigma + {\cal L}_v \sigma = 0.
\end{align}
Hence, we identify the 3-form field strength variables with fluid variables in the following way:  $H=\sigma$ and $H^i = \sigma v^i$.

\section{Action Principle for 3+1 dimensional Euler equations}

In this section we propose two possible actions to describe an incompressible fluid in 3+1 dimensions.  In the first case we describe the ``vorticity sector" of the theory with a 1-form $A_\mu$ and the associated magnetic (\ref{Bfield}) and electric fields (\ref{Efield1}).  This leads to an explicit BF term in the action, as described above.  However, as we will see below, the resulting field equations restrict the ``helicity sector".  To remedy this issue, we re-express the vorticity sector in terms of Clebsch variables and an auxiliary gauge connection, generalizing the construction in \cite{Tong:2022gpg}.

First, following the 2+1 dimensional action (\ref{2daction}), we propose the following action for an incompressible fluid in 3+1 dimensions:
\begin{align}
S_{\rm test} = \int dt d^3x ~ \left(\frac{H_i H^i}{2H} - pH + \frac{1}{2} \epsilon^{\mu \nu \rho \sigma}  P_{\mu \nu} F_{\rho \sigma} \right).
\end{align}
This action is invariant under rotations and space-time translations, but not Lorentz boosts.  The action is even under time reversal, but changes by an overall sign under parity.

To test this action,  we vary with respect to $P_{\mu \nu}$.  Note that we define $H_{\mu \nu \rho} = \partial_{[\mu} P_{\nu \rho]}$.  Then it follows that
\begin{align}
\delta H &= \epsilon^{ijk} \partial_i (\delta P_{jk}) \nonumber \\
\delta H^i &= \epsilon^{ijk} \left(\partial_t (\delta P_{jk}) + \partial_k (\delta P_{j0}) + \partial_j (\delta P_{0k}) \right)
\end{align}
First, we consider variations with respect to $P_{0i}$.  This yields the constraint equation
\begin{align}
-\epsilon^{ijk} \partial_j \left(\frac{H_i}{H} \right) - B^i = 0
\end{align}
In terms of velocity $v^i = H^i/H$.  As before, we assume that $\vec{v}$ and $\vec{\omega}$ are a priori independent variables.  The constraint equation connects them via the expected relation that $\vec{\omega}$ is the curl of the velocity
\begin{align}
\vec{\nabla} \times \vec{v} = \vec{B} = \vec{\omega} \label{vorticitycurl}.
\end{align}

Next we vary with respect to $P_{ij}$.  Peeling off an overall factor of $\epsilon^{ijk}$ yields the following field equation in terms of the fluid variables
\begin{align}
\partial_t v_k + \partial_k (v^2/2) + \partial_k p + E_k = 0.
\end{align}
Inserting the form of the electric field (\ref{Efield1}), imposing the constraint equation (\ref{vorticitycurl}) and using the vector calculus identity
\begin{align}
\partial_i (v^2/2) = v^j \partial_j v_i + \vec{v} \times \vec{\omega},
\end{align}
yields the incompressible Euler equation.

We must also vary the 1-form connection $A_\mu$ in the $F_{\mu \nu}$ part of the topological term.  This gives field equations that require the local helicity to vanish
\begin{align}
H = H^i = 0,
\end{align}
which would restrict us to flows with zero local helicity.  In terms of the helicity density $h$
\begin{align}
h = \vec{v}\cdot \vec{\omega} = \vec{\omega} \cdot \vec{\nabla} \ell.
\end{align}

One simple way to allow for generic non-zero $H=\sigma$ in this framework is to couple the theory to an external current $J^\mu_{\rm ext}$, such that the total action is
\begin{align}
S_{\rm total} = \int d^4 x \left(\frac{H_i H^i}{2H} - pH - \frac{1}{2} \epsilon^{\mu \nu \rho \sigma}  P_{\mu \nu} F_{\rho \sigma} + J_{\rm ext}^\mu A_\mu \right) \label{3daction}
\end{align}
Because the current is conserved, the action remains gauge invariant.  In this case the field equations tie the local helicity density to the electric charge density, eg.
\begin{align}
H &= \rho \nonumber \\
H^i &= \rho v^i.
\end{align}
If there are a set of particles with charge $e$, located at positions $x_a$, then $\rho(\vec{x},t) = \sum_{a=1}^{N} e \delta(\vec{x}-\vec{x}_a)$ and
\begin{align}
H(\vec{x},t) = \sum_{a=1}^{N} e \delta(\vec{x}-\vec{x}_a(t)).
\end{align}

\subsection{Clebsch BF action}

To address the restriction on the local helicity in the first action, we now consider the following action again in terms of $P_{\mu \nu}$, but now with the field strength (or alternatively, the 2-form vorticity current) depending on scalars $\alpha$ and $\beta$ as follows
\begin{align}
S_{\rm clebsch} = \int dt d^3x ~ \left(\frac{H_i H^i}{2H} - pH +  \frac{1}{2} \epsilon^{\mu \nu \rho \sigma}  P_{\mu \nu} \partial_\rho \alpha \partial_\sigma \beta \right)  \label{Clebschaction}.
\end{align}
The scalars $\alpha$ and $\beta$ can be thought of as being charged under $P_{\mu \nu}$.  Furthermore, one can define an auxiliary connection $\tilde{A}_\mu = \alpha \partial_\mu \beta + \partial_\mu \chi$, which is a Clebsch decomposition of the field.  Hence the last term in the action is a ``Clebsch BF" term
\begin{align}
S_{\rm clebsch} = \int dt d^3x ~ \left(\frac{H_i H^i}{2H} - pH + \frac{1}{2} \epsilon^{\mu \nu \rho \sigma}  P_{\mu \nu} \tilde{F}_{\rho \sigma} \right).
\end{align}
Variation of this action with respect to $\alpha$ and $\beta$ yields
\begin{align}
H^0 \partial_t \alpha + H^i \partial_i \alpha = 0 \nonumber \\
H^0 \partial_t \beta + H^i \partial_i \beta = 0.
\end{align}
These equations express the familiar fact that the Clebsch scalars are material invariants.  Varying the action with respect to $P_{0i}$ gives
\begin{align}
\vec{\nabla} \times \vec{v} = \vec{\nabla} \alpha \times \vec{\nabla} \beta,
\end{align}
which the correct relationship between the curl of the velocity/vorticity vector and the Clebsch scalars.  Finally, varying with respect to $P_{ij}$ gives the same terms as before, but now with the Clebsch contribution $\epsilon^{ijk} (\partial_t \alpha \partial_k \beta - \partial_k \alpha \partial_t \beta)$.  This term can be re-expressed as the $\vec{v} \times \vec{\omega}$ term and the resulting field equation is again the Euler equation.  Therefore, the Clebsch BF action has the advantage of reproducing the Euler equations without any constraints on the local helicity.

\section{Boundary terms, gauge invariance, and edge modes}

We now consider the two gauge theory actions in the presence of a boundary, which leads to interesting new physical effects.  In the dual fluid, this setting corresponds to a flow with a boundary.  We first note that the variation of the actions has the following generic form
\begin{align}
 \delta S_{\rm total} = \int dt d^3 x ~ \left(\frac{\delta L}{\delta \psi} \delta \psi + \partial_\mu S^\mu \right),
\end{align}
where $L$ is the Lagrangian, $\psi = (A_\mu, P_{\mu \nu})$ or $(\alpha, \beta, P_{\mu \nu})$.  Setting the first term on the right-hand side equal to zero imposes the field equations, but to have a well-defined variational principle, the surface terms in $S^\mu$ must also vanish.  For (\ref{3daction}) and a spatial boundary, the boundary terms are
\begin{align}
\delta S_{\rm total, bdry} = \int dt d^2 x ~ \left(2 n_k v_j \epsilon^{ijk} \delta P_{0i} - (p+\frac{1}{2}v^2) n_i \epsilon^{ijk} \delta P_{jk} + 2 n_i \epsilon^{ijk} P_{0j} \delta A_k + n_i \epsilon^{ijk} P_{jk} \delta A_0   \right),
\end{align}
where $n^i$ is the normal to the boundary surface.  In the case where the boundary is a plane at $x=0$, then the boundary terms become
\begin{align}
\delta S_{\rm total, bdry} = \int dt dy dz \left(2 \epsilon^{ab} v_a \delta P_{0b} - (p+\frac{1}{2}v^2) \epsilon^{ab} \delta P_{ab} + 2 \epsilon^{ab} P_{0a} \delta A_b + \epsilon^{ab} P_{ab} \delta A_0   \right),
\end{align}
where $x^a = (y,z)$.  One way to set these terms to zero is to fix $P_{0a} = P_{xy} = 0$, which would leave the 1-form gauge field free on the boundary.  More generically one could fix the boundary values of the 2-form potential to be constants $k_a$ and $l$:
\begin{align}
P_{0a} &= k_a \nonumber \\
P_{xy} &= l.
\end{align}
Then fixing $2 \epsilon^{ab} k_a A_b + l A_0 = 0$ on the boundary eliminates the remaining boundary contribution to the action.

For the Clebsch BF action (\ref{Clebschaction}) the boundary terms are the same, except the contribution from variations of $A_\mu$ are replaced by variations of the Clebsch scalars
\begin{align}
 \int dt d^2 x \left(-2 n_i \epsilon^{ijk} P_{0j} \partial_k \beta~ \delta \alpha - n_i \epsilon^{ijk} P_{jk} \partial_t \beta~ \delta \alpha + 2 n_i \epsilon^{ijk} P_{0j} \partial_k \alpha ~\delta \beta + n_i \epsilon^{ijk} P_{jk} \partial_t \alpha~ \delta \beta \right).
\end{align}
Setting $P_{0a} = P_{xy} = 0$ will again remove the boundary contribution, at the expense of leaving the Clebsch scalars free on the boundary.  As in the other theory, could also fix the boundary values of $P_{\mu \nu}$ to be constants on the boundary and arrange for a combination of $\alpha$ and $\beta$ to be fixed.

Another issue that we alluded to earlier is that both theories are only gauge invariant up to a total derivative.  Therefore, gauge invariance must be reconsidered in the presence of a boundary.  Under the 2-form gauge transformation $P_{\mu \nu} \rightarrow P_{\mu \nu} + \partial_{[\mu} \lambda_{\nu]}$ the action transforms as
\begin{align}
S_{\rm total} \rightarrow S_{\rm total} + \int dt d^2 x \left(n_i B^i \lambda_0 + n_i \epsilon^{ijk} \lambda_j E_k\right).
\end{align}
The gauge variation of $S_{\rm clebsch}$ is the same, with $\vec{B}$ and $\vec{E}$ expressed in terms of Clebsch variables.  Hence, the gauge variation of both actions is zero when conductor-like boundary conditions are present, ie when the normal component of the magnetic field and the transverse components of the electric field are zero at the boundary.  In fluid variables this implies
\begin{align}
\vec{\omega} \cdot \vec{n} = 0 \nonumber \\
(\vec{v} \times \vec{\omega})\cdot \vec{t} = 0.
\end{align}
These conditions require both the vorticity vector and the flow velocity to be tangent to the boundary surface. The boundary conditions we imposed above to eliminate the boundary term in the action are not enough to force the normal component of the magnetic field and the transverse electric field to vanish.  If these conductor-like conditions do not hold, one could still save gauge invariance by requiring that 2-form gauge parameter satisfy $\lambda_0 = \lambda_a = 0$ on the boundary.  In addition, in the first theory the 1-form gauge transformations also must be restricted to the subset of transformations that preserve the boundary condition on $A_0$ and $A_a$.   All these requirements reduce the amount of gauge freedom of the theories.

Reducing the gauge redundancies of a theory means that degrees of freedom that were previously pure gauge become physical on the boundary surface.  Alternatively, one could introduce new fields on the boundary designed to restore the gauge invariance of the combined bulk plus boundary system.  Either approach leads to the existence of edge modes, as was first noticed in Chern-Simons theories, where the boundary theory is a 1+1 dimensional chiral scalar field \cite{Elitzur:1989nr,Wen:1992vi,Balachandran:1991dw}.  For BF theories a similar analysis was performed first in the context of field theory \cite{Balachandran:1992qg,Balachandran:1993wj} and then later in the study of edge states of 3d topological insulators \cite{Hansson:2004wca,Cho:2010rk}.

To study the edge states of our gauge theories, we first consider their symmetries, which can be obtained by computing the Noether charges associated with the gauge symmetries.  In the absence of boundaries, the Noether charges for local symmetries are zero on-shell, but when a boundary is present, the charges become non-trivial integrals over the boundary surface, associated with gauge transformations that act as physical transformations on boundary degrees of freedom.  Using the textbook formula for the Noether current of a local gauge symmetry, parametrized by $\lambda$, $Q = \int j^0 d^3 x$, yields generically
\begin{align}
Q(\lambda) = \int d^3 x ~ C(\lambda) + \oint d^2 x ~ q(\lambda)
\end{align}
$C$ represent the constraint equations and $q$ is a boundary charge aspect.  For (\ref{3daction}) the charge associated with the 2-form symmetry is
\begin{align}
Q^{(2)}_{\lambda_i} = \int d^3 x ~ \left( - \epsilon^{ijk} \partial_j \left(\frac{H_i}{H}\right) + \epsilon^{ijk} \partial_j \left(\frac{H_i}{H}\right) \lambda_k - \lambda_i B^i \right)
\end{align}
Imposing the constraint equation and Stokes theorem gives
\begin{align}
Q^{(2)}_{\lambda_i} = \oint d^2 x ~n_i \epsilon^{ijk} v_j \lambda_k. \label{charge2}
\end{align}
Following the same procedure for the 1-form symmetry gives
\begin{align}
Q^{(1)}_\lambda = \oint d^2 x ~ \lambda n_k \epsilon^{kij} P_{ij}.
\end{align}
Note that when $\lambda = constant$, $Q^{(1)}$ is equal to the integrated helicity inside the region of the flow
\begin{align}
Q^{(1)}_c = \int d^3 x ~ \sigma,
\end{align}
reflecting the material invariance of this quantity.  The infinite number of charges for generic time independent functions $\lambda$ are a generalization of the Kelvin theorem for helicity. Here the $\lambda$ represent ``improper" gauge transformations.  The fluid interpretation of $Q^{(2)}$ is less clear; it appears to be a new type of conserved quantity.

For the Clebsch BF action (\ref{Clebschaction}) the Noether charge $Q^{(2)}$ associated with the 2-form gauge symmetry is the same, ie (\ref{charge2}).  The 1-form gauge symmetry is not present, instead there is an invariance under area preserving maps in the 2d space of Clebsch scalars (ie, symplectomorphisms), which preserve the vorticity 2-form current.  For $c_a = (\alpha, \beta)$
\begin{align}
c_a = M_a (c),~~  \frac{\partial(M_1, M_2)}{\partial(\alpha, \beta)}= 1.
\end{align}
The gauge connection acts as the canonical 1-form $p dq$, while the vorticity is the canonical 2-form $dp \wedge dq$.  It would be interesting to flesh out the precise relationship between the 1-form gauge transformations on $A_\mu \leftrightarrow p dq$ and the symplectomorphisms.

Finally, the algebra of the Noether charges can be determined using the Poisson bracket, e.g.
\begin{align}
\{Q_{\lambda_1}, Q_{\lambda_2}\} = \delta_{\lambda_2} Q_{\lambda_1}.
\end{align}
The 1-form and 2-form charges commute with each other due to the gauge invariance of their charge aspects
\begin{align}
\{Q^{(1)}_{\lambda_1}, Q^{(1)}_{\lambda_2}\} &= 0 \nonumber \\
\{Q^{(2)}_{\lambda_{i,1}}, Q^{(2)}_{\lambda_{j,2}}\} &= 0.
\end{align}
However, in the explicit BF theory, due to the 2-form gauge dependence of $Q^{(1)}_\lambda$, it follows that
\begin{align}
\{ Q^{(1)}_{\lambda}, Q^{(2)}_{\lambda_{j}}\} = \oint d^2 x ~ \lambda n_k \epsilon^{kij} \partial_i \lambda_j.
\end{align}
The same charge algebra was found in \cite{Balachandran:1992qg,Balachandran:1993wj} for the case of pure topological BF theory.  Like Chern-Simons theories, non-topological terms in the action do not modify the charge algebra \cite{Park:1998yw}.

Hence, we assume that the dynamics of the edge modes in (\ref{3daction}) are controlled by topological BF term.  The simplest procedure is to fix a gauge where $A_0 = P_{0i} = 0$.  Solving the constraint equations of the pure BF theory yields $A_i = \partial_i \phi$ and $P_{ij} = \partial_{[i} \rho_{j]}$. Normally these are pure gauge degrees of freedom, but substituting this ansatz into the action yields as non-trivial boundary action \cite{Balachandran:1993wj,Amoretti:2012hs}
\begin{align}
S_{\rm edge, BF} = \int dt d^2 x ~\left( \epsilon^{ij} \partial_i \rho_j (\partial_t \phi) + \partial_i \phi \epsilon^{ij} \partial_t \rho_j \right).
\end{align}
It has been argued that this edge dynamics appears as the self-dual sector of a theory of a free scalar plus a Maxwell field on the boundary. In terms of the fluid picture, our results suggest there are edge modes appearing when the 3d flow is subject to boundaries.  However, further work is needed to understand the role of these modes in the fluid picture.

\section{Discussion}

In this paper we have constructed gauge theories with a BF term that describe the incompressible Euler equations in 3+1 dimensions. Both theories contain a 2-form gauge sector associated with local helicity.  However, in one theory, the BF term is explicit; there is a 1-form gauge invariance and vorticity are expressed in terms of the associated magnetic field. The drawback with this approach is that local helicity is constrained by the field equations to vanish.  In the other version, the BF term involves Clebsch scalars, which parametrize the vorticity current. In this case, local helicity is unconstrained. We argue that the Clebsch BF action is likely the proper description of 3d flows in general, coupling the local helicity described by the 2-form potential with the vorticity described by the Clebsch scalars.  However, our results suggest that incompressible flows with zero local helicity may have a purely topological description in terms of BF theory.  Finally, we also studied the theories in the presence of a boundary, computing boundary Noether charges and using past results on BF type theories to postulate that there are physical degrees of freedom living on the boundary surface (ie, edge modes).

One obvious area of investigation is whether the gauge theory picture can be extended to include a viscosity term, thereby reproducing the Navier-Stokes equations.  A viscous term contributes to the helicity transport equation as $\nu \nabla^2 \sigma$, which means that the Bianchi identity equation must be modified.  This suggests that viscosity may break the gauge invariance of our theory.

On the other hand, the viscous terms in the Navier-Stokes equation are associated with the diffusion of vorticity and momentum, whose effect can be modeled by introducing a stochastic noise term into the equation for the map between Eulerian and Lagrangian variables \cite{Gomes, ConstantinIyer}. Physically, the noise introduces a randomness in the Lagrangian paths of the fluid parcels.  The average of the stochastic system satisfies the Navier-Stokes equation.

Eqns. (\ref{mapgauge}) and (\ref{mapgauge2}) relating the Euler/Lagrangian maps and the gauge fields indicate that treating particle paths as random also introduces randomness in the gauge field.  It would be interesting to see if a functional integral version of our theory, which viscosity playing the role of $\hbar$, can be used to describe the Navier-Stokes system.  In terms of a partition function
\begin{align}
Z = \int D[A] D[P] e^{- S_{\rm gauge}/\nu}.
\end{align}
The idea is that expectation values of gauge invariant fluid observables would obey the Navier-Stokes equations.  In addition, in the stochastic setting one can prove that a generalized  version of the Kelvin theorem holds \cite{EyinkAction}, meaning that that currents are conserved in an averaged sense, with Ward-like identities, eg
\begin{align}
\langle \partial_\mu J^\mu \rangle = 0 \nonumber \\
\langle \partial_\mu J^{\mu \nu} \rangle = 0.
\end{align}
A related issue is the role that gauge invariance plays in turbulent states, for example in the infinite Reynolds number limit where $\nu \rightarrow 0$.  In this limit there are singularities in the flow associated with vortex sheets and lines.

It would also be interesting to elucidate the role of the edge modes predicted by our theory in incompressible fluid flows with a boundary.  For example, in 2d, we argued that lines of zero vorticity are natural boundaries in flows.  The changing of the sign of the vorticity leads to a spontaneous breaking of time reversal symmetry and the presence of chiral modes propagating on the boundary surface \cite{Eling:2023iyx, Eling}.  In 3d, zero vorticity lines or surfaces may have similar chiral modes.  However, it may be the case that non-trivial edge modes exist even in cases where time reversal symmetry is unbroken.

\end{document}